# Knowledge Management for Cloud Computing Field

Mahdi Fahmideh, University of Southern Queensland

Jun Yan, University of Wollongong

Jun Shen, University of Wollongong

Aakash Ahmad, University of Ha'il, Saudi Arabia

Davoud Mougouei, University of Southern Queensland, Australia

Anup Shrestha, University of Southern Queensland, Australia

**Abstract.** Migration legacy systems to cloud platforms is a knowledge intensive process. There is an ever increasing body of knowledge reporting empirical scenarios of successful and problematic cloud migration. Reusing this body of knowledge, dispersed and fragmented over the academic/multi-vocal literature, has practical values to mitigate costly risks and pitfalls in further projects of legacy-to-cloud and cloud-to-cloud migration. In line with this, knowledge management systems/platforms pertinent to cloud migration are a prime prerequisite and a strategic imperative for an organization. We have conducted a qualitative exploratory study to understand the benefits and challenges of developing Knowledge Management Systems (KMS) for cloud migration in real trials. Whilst our prototype system demonstration supported the importance and benefits of developing Cloud Migration KMS (CM-KMS), our semi-structured industry interview study with 11 participants highlighted challenging impediments against developing this class of KMS. As a result, this study proposes nine significant challenges that cause the abandon of the design and maintenance of CM-KMS, including continuous changes and updates, integration of knowledge, knowledge granularity, preservation of context, automation, deconstruction of traditional knowledge, dependency on experts, hybrid knowledge of both vendor-specific and vendor-neutral cloud platforms, and parsimony. Our results inform cloud architects to pay attention to adopt CM-KMS for the legacy-to-cloud migration in their organizations.

## 1 Introduction

Momentum is growing to migrate legacy software systems to cloud services in different possibilities such as re-hosting, lift and shift, re-build, or switching between multiple cloud platforms with different service delivery models, i.e. IaaS (infrastructure as a service), SaaS (software as a service), and PaaS (Platform as a service) [1]. It is commonly acknowledged that cloud services positively contribute to the operational performance of legacy systems in terms of scalability, storage capacity, and processing speed. The process of moving monolithic legacy software systems from on-premise infrastructure to cloud platforms and between multiple cloud platforms, on the other hand, is knowledge extensive and demands deep knowledge in many technical and non-technical aspects such as legacy code refactoring, interoperability across different cloud platforms, department downsizing, vendor lock-in, and many others [2],[3],[4]. Cloud system architects (architects for short) need to know all potential risks and drawbacks before and during the transition process, and apply certain countermeasures to mitigate them beforehand [1],[2]. Moreover, as cloud services are quite unpredictable, both in terms of software execution and updates of underlying platforms, a key concern of organizations at the early stage of cloud migration is to ensure that if the promising benefits of cloud services will be achievable in practical settings [5]. This may raise primary concerns such as "do cloud platforms positively contribute to the scalability of our legacy systems or conversely cause adverse effects?" or "what unexpected issues may occur in our systems after deployment on cloud servers?" These questions can be answered after actual trial once systems are in operation in the cloud, but, this might be a costly exercise to rectify if issues of cloud services occur after the transition.

To address these queries, the *knowledge reuse approach* [6] suggests that if errors, risks, or successes related to system development endeavors are continuously collected, annotated, and explicated, they help avoid repeating pitfalls, hence improving the quality of systems at a lower cost. For example, if the architect wants to know how to tackle the risk of *legacy system personal data remanence on cloud servers*, as it is a commonly reported data privacy issue in cloud migration scenarios [1],[7], she can look into KMS to find how one's peer has already addressed this issue in similar scenarios. Such KMSs facilitate knowledge reuse, communication across development teams, and teaching disciplinary knowledge to newcomers in the cloud computing field [3],[8].

While emerged KMS, specific to cloud migration, has attracted much research interests in this area, the way forward for developing and using KMS related to cloud migration is unclear. Moreover, cloud architects need to be able to relate KMS to the challenges they face. In some cases, cloud architects are only marginally convinced that KMS provides them with any real benefit. They may also reluctant to share the challenges or failures of KMS due to their organization's policies and privacy. Whereas successful KMS utilizations in cloud migration projects can be outstanding endeavors that help learn and perform a safe migration of legacy systems to cloud (e.g., [5],[7],[16],[18]), identification of adoption challenges play an equal role in the successful adoption of KMS initiatives. These challenges, causing the abandon of KMS projects, indeed demonstrate the dark side of this process for cloud architects. Ignorance of these challenges may jeopardize the KMS projects in cloud migration context and cause the wastage of resources.

Attempting to underline the important challenges in design and adoption of KMS, we applied Design Science Research (DSR) methodology [10] and developed a prototype artefact called *Cloud-Migration Knowledge Management System (CM-KMS)* as a testbed instance. It enables system architects to reuse cloud migration evidential knowledge and rationalize requirements analysis prior to running into the implementation phase. Our CM-KMS draws on the evidential data on obstacles/risks against the common system quality goals as well as the corresponding resolution tactics/countermeasures to tackle these obstacles in the context of cloud migration. The evidential datasets are sourced from the academic and multi-vocal literature [9]. We demonstrate the application of CM-KMS in an exemplar scenario of migrating a legacy hyper-connected manufacturing collaboration system (HMCS) to data analytics platforms. The scenario verifies the benefits of CM-KMS's evidential data that are incorporated into the requirements analysis activities during HMCS migration. Next, we conducted an industry interview study with 11 selected participants from our industry partners and academia in Europe, Australia, and Asia. From the semi-structured interviews, we extracted contemporary challenges that lead architects to abandon or avoid the use of CM-KMS.

The contribution of our study to theory and practice of cloud migration literature is the laying out challenges for taking action in future KMS development endeavors in the emerging closely-related topic such as migrating legacy software systems to microservice architecture, Internet of Things, and blockchain smart contracts.

The rest of this paper is organized as follows. Section 2 presents the theoretical background of this research and reviewing the existing works in the literature. This is followed by Section 3 where we discuss DSR methodology that we applied to conduct this study. This continues with an illustrative application scenario of CM-KMS prototype in Section 4 and reporting the results of the interview study with participants in Section 5. Finally, the limitations of this research and future research directions are discussed in Section 6.

## 2      Research Background

The Oxford Dictionary defines *knowledge* as specific information, facts, and intelligence about something. In this research, the term knowledge is contextualized to describe the obstacles causing the failure of commonly intended cloud migration goals as well as high-level countermeasure/resolution tactics as the means to tackle these obstacles. Stenmark refers to the main types of knowledge fragments namely, *tacit* and *explicit* [14]. Tacit knowledge is the one that cannot be explicitly expressed, though it guides the behavior of the human. Explicit knowledge, on contrary, which is the focus of this research, can be explained in the form of reports, books, talks, or other formal/ informal communication.

Whilst achieving efficient knowledge management through the adoption of cloud computing technology has received significant attention in the literature [8], the opposite aspect of the matter, i.e., challenges that cause to demise the development and maintenance of KMS in cloud migration project, has been less explored. The existing research on applying knowledge reuse is mainly focused on the implementation and coding stages of cloud-enablement. For instance, Cloud Life Cycle is a repository containing historical information on QoS cloud service market increasing the accuracy of service selection [15]. More specifically, Zimmermann in [7] suggests reusing and sharing recurring decision logs related to re-architecting systems to the cloud. It provides a catalogue of reusable architectural refactoring templates that are confined to improving the scalability of cloud solution architecture. Using software patterns, aka *"recurrent solutions to frequent problems"*, either generic (e.g., Cloud Computing Patterns [16], MODAClouds [17]) or vendor-specific (e.g., AWS Cloud Design patterns (CDP) [18],

Microsoft [19], and IBM [20]) are a means to improve the performance of cloud-enabled architecture design. The fine granular patterns and authoring toolkits are technical-oriented and related to implementation stages such as virtualization, workload bursting, elastic environment, multi-tenancy, and cloud deployment models [18]. Fahmideh et al. [5] emphasize the value of capturing and reusing the evidential knowledge from cloud migration scenarios that are based on rigorous and reproducible research methods. However, their work suffers from formal evaluation by system architects in real-world scenarios. Unlike our work in this research, Fahmideh et al. [5] do not provide recommendations guiding the design of KMSs for the cloud computing field as a class of design problem. We provide empirical industry-focused research with a prototype system and in-depth interviews, where we combine different perspectives in order to find out which challenges occur in the development and maintenance of CM-KMS. We leverage the ideas in general KMS design principles [6],[11],[12],[13] and the software architectural knowledge management community [11] to create our cloud-specific KMS which relies on evidential knowledge of cloud computing migration.

Prior studies regarding KMS in cloud computing field focus either on developing KMS or on available solutions in science and commercial tools. In comparison, our work takes a holistic view and reports empirical industry-focused research with in-depth interviews, in which we combine different perspectives in order to find out which emerging challenges occur in developing and adopting KMS or are unresolved attuned with migrating legacy systems to cloud platforms. This is necessary to gauge the adoption of new KMS technologies in practice. The nine identified challenges in our study can be incorporated into developing new KMS or adopting existing ones available in marketplace in cloud migration projects, thus helping cloud architects in organizations to avoid issues leading to the demise of KMS in cloud migration lifecycle.

## 3    Research Design

Our research objective can be defined as follows: Analysis of the contemporary challenges of developing and adopting KMS for the purpose of reducing risks in migration legacy systems to cloud platforms. To address this research query and find out how the class of CM-KMS works in practice, we applied the DSR methodology [10]. DSR has a decade of an old tradition in information systems and software engineering research. One depicted and well-known model of DSR developed by Peffers and Tuunanen [22] defines phases of design, demonstrate, and evaluate as described in the following.

### 3.1    Design iterations

A DSR artefact design should be based on prior kernel approaches in a given field of research to provide input and justifiable knowledge to the design cycles guiding the creation of the artefact [23]. The development of CM-KMS has relied on the existing interrelated design requirements (DRs) and design principles (DPs) that are grounded in KMS design in the software engineering literature [6],[9],[11],[12],[13], [24] and specialized for the context of this research. Table 1 shows DRs that were served as

5high-level guidelines and anchored during the design, demonstration, and evaluation phases. For instance, in addressing DR1 (soundness and relevance), DP1 was defined based on the listed challenges related to migrating legacy systems to cloud platforms [1]. Similarly, DP2 refers to providing empirical support for knowledge fragments and has been informed by [9]. DP4 is informed by [12] recommending a formal structure to represent the knowledge fragments.

Table 1. Design requirements and principles incorporated in our DSR phases adopted from [6],[9], [11],[12],[13]

| |
|---|
| **DR1. Soundness and relevance** |
| DP1. Identify knowledge fragments, i.e., obstacles and countermeasures, specific to cloud migrating requirements and perceivable and soundness by experts. |
| DP2. Ensure the evidential support for the knowledge fragments (e.g. reported research method, migration scenario context) to give reliability. |
| **DR2. Classification and integrity** |
| DP3. Consider variant of service delivery models (e.g., IaaS, SaaS, PaaS) and migration types, e.g., re-hosting/lift-shift, re-build. |
| DP4. Ensure knowledge fragments are well-structured and in relation together to facilitate further knowledge base updates. |
| DP5. Consider integrity with other knowledge sharing systems. |
| **DR3. Granularity and hierarchy** |
| DP6. Determine the level of details to be captured in knowledge base. |
| DP7. Define a representation showing knowledge storing in knowledge base. |
| **DR4. Decision support** |
| DP8. Consider the context for knowledge fragments (e.g. the choice cloud platform and service delivery model) and related situational factors to overcome one-size-fits-all decision making issues. |

The abovementioned design principles were incorporated into our Systematic Literature Review (SLR) [9] to compile CM-KMS knowledge repository datasets. The two objectives of SLR were to identify *(i) evidential obstacles against common system quality goals that have been reported in the academic and multi-vocal literature of migrating legacy systems to cloud* and *(ii) evidential countermeasure to negate these obstacles*. We followed the guidelines from [25] and strictly performed the recommended steps of SLR [9] to identify published empirical results of legacy to cloud migration in academic and multi-vocal literature accounting challenges in migrating systems to cloud platforms. We manually performed the keyword-based search with *migration*, *legacy systems*, and *cloud computing* as the main terms combined with their synonyms via AND/OR operators as shown in Table 2.

Table 2. An excerpt of search strings (SS)

| |
|---|
| SS1: "Migration" OR "Cloud adoption" OR "Cloud migration" OR "Migration to cloud" OR "Legacy to cloud migration" OR "Legacy migration to cloud" AND [SS2 OR SS3 OR SS4 OR SS5 OR SS6] |
| SS2: "IaaS risks" OR "IaaS challenges" OR "IaaS challenges" OR "IaaS adoption" OR "IaaS benefits" |
| SS3: "PaaS risks" OR "PaaS challenges" OR "PaaS issues" OR "PaaS adoption" OR "PaaS benefits" |
| SS4: "SaaS risks" OR "SaaS challenges" OR "SaaS issues" OR "SaaS adoption" OR "SaaS benefits" |

SS5: "Monolith application" OR "Legacy code" OR "Legacy system" OR "Existing system" OR "Legacy component" OR "Legacy software" OR "Legacy application" "On-premise application" OR "Monolithic system" OR "Existing software" OR "Pre-existing software" OR "Legacy information system" OR "Legacy program" OR "Pre-existing assets" OR "Legacy architecture" OR "Legacy asset"

SS6: "Reengineering" OR "Legacy system reengineering" OR "System reengineering"

The criteria for study selection were, for example, (i) the availability of a clear description of scenario of legacy system migration to the cloud and (ii) reporting cloud-specific issues in migrating legacy software to cloud platforms. The search was performed over the scientific digital libraries including IEEE Explore, ACM Digital Library, SpringerLink, ScienceDirect, Wiley InterScience, ISI Web of Knowledge, and Google Scholar. The leading IS and Software Engineering journal/conference proceedings were particularly selected including IEEE Transactions on Software Engineering/Cloud Computing (TSE and TCC), JSS, IST, TOSEM, the senior basket of eight Information Systems journals, SIG recommended journals by AIS. We reviewed studies published between timespan 2007 and late 2020. Backward and forward *snowballing technique* [26] was used to avoid missing important studies. In total, 115 studies were identified after applying the inclusion and exclusion criteria. The following publication metadata were extracted from each identified paper: author, title, publication channel, year, publisher, country, type, study aim, validation techniques, and findings. We found 11 common system quality goals as the basis to derive the obstacles and resolution tactics from the identified papers. That is, for each goal, we checked the paper of any reported conditions causing the quality goal failure. The full text of each paper was thoroughly reviewed and text segments related to two objectives of SLR were extracted and coded under the datasets of obstacles and resolution tactics within several iterative processes of coding. This literature review allowed us to create 30-pages document of evidential data, publically available at *https://tinyurl.com/KSP-CloudMigration*, that was then stored in the repository of CM-KMS prototype system.

### 3.2 Quality assessment and statistics

As a yardstick to select and extract evidential data from the identified studies, we took into account that the number of citations that the study had received is a tentative indicator of its significance [11]. That is, the higher number of citations, the higher the level of agreement and the more canonical and significance of the study in opinions of the discipline's scholars. During the design, DPs were interleaved with the compilation of the datasets. For instance, we organized the datasets based on the conceptual model shown in Fig. 1, to support DP5.

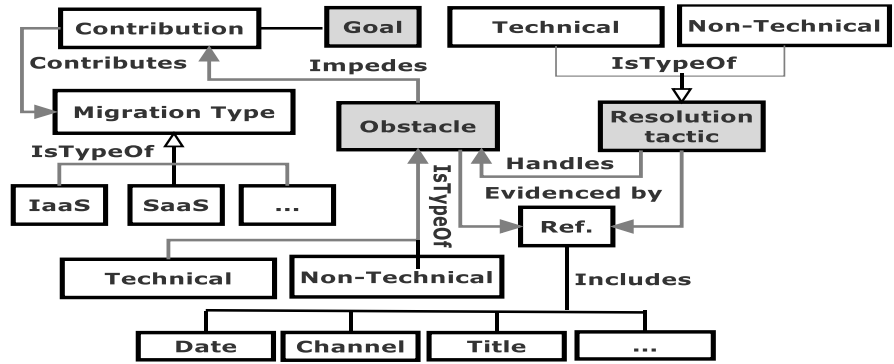

Fig 1. The conceptual model for the representation of knowledge repository datasets

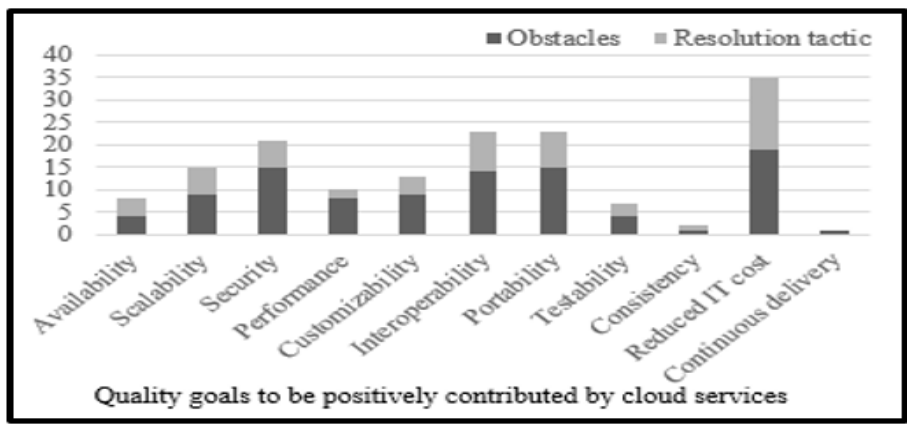

Fig. 2. frequency of accounted obstacles against quality goals derived from the literature

We tried providing meta-data for each item in the datasets such as the relations of an obstacle to a migration type and resolution tactics in line with (DP10) to provide a meaningful connection between the evidential data and the real-world cloud migration scenarios. We came up with the datasets: (i) *obstacles/risks* including 70 common probable technical and non-technical situations hindering legacy systems benefit from cloud services; and (ii) *resolution tactic* comprising 56 cloud platform-agnostic mechanisms to handle these obstacles. Fig. 2 shows the occurrence rate of obstacles against the common system quality factors. For instance, whilst *reduced legacy system cost* is a common goal in migrating legacy systems to the cloud, there are 19 accounted obstacles that impede achieving this goal. This includes, for example, *extra security cost* (reported by studies S6, S26, S28), *high cancellation fees* (reported by S33), *learning curve* (reported by S43), *licensing issue* (reported by S45, S55), *department downsizing* (S44, S56), and *variable price of cloud resources* (S112) (Table 3). *Security* and *portability* are the second most vulnerable goals that are threatened by 15 obstacles. We stored the datasets in CM-KMS's prototype repository. The complete set of datasets in a text format is publically accessible at *https://tinyurl.com/KSP-CloudMigration*.

Table 3. An excerpt of knowledge fragment related to probable obstacles against the quality goal along resolution tactics (source: *https://tinyurl.com/KSP-CloudMigration* )

| Quality goal | Definition | Source |
|---|---|---|
| Reduced IT cost | Lower expense for infrastructure procuring, data storages, system updates, maintenance, and staff. | [S2], [S3], [S4], [S5], [S35], [S36], and [S37]. |
| **Obstacle** | **Definition** | **Source** |
| Department downsizing | The maintenance team of legacy systems may become downsize as their responsibilities are outsourced to cloud providers. | [S44], [S56] |
| **Resolution tactic** | **Definition** | **Source** |
| Analyze migration feasibility | Perform a feasibility analysis to evaluate the benefits and the consequences of moving legacies to the cloud and its impact on organization structure, staff's roles, and legacies. | [S73], [S74] |

## 4  Demonstration: an application example

We specialized HMCS project [27] to illustrate how the evidential knowledge in CM-KMS prototype system is helpful to inform the architect of requirements that should be addressed in a legacy system cloud enablement scenario. ELT (Extract-Load-Transform) was the key component of HMCS architecture collecting and blending data from different data sources through a middleware layer defining business logic and transformation rules for data mapping between these sources. ETL was collecting data from two sources: (i) various equipment such as sensors and smart tags installed in the product line of the car manufacture that monitor events that may raise urgent issues and (ii) online client conversations about the car products such as recent purchase experiences, warranty claims, repair orders, and service reports that are posted on Twitter, Facebook, and blogs. The heterogeneity of data sources and the large volume of incoming data has made ETL a bottleneck (as-is concern) that would most probably remain unresolved even after adding new local servers. To tackle this issue, the architects suggested to developers to use cloud-based Microsoft Azure Apache Hadoop platform as an appropriate solution because it could provide a rich set of tools, programming, and scripting frameworks increasing the performance of complex ETL's functions. What-if questions asked by the architect via CM-KMS step-by-step procedure and interactive forms were (Fig. 3): (i) how ETL quality goals, e.g. availability or security, might be generally influenced if Hadoop platform is utilized? (ii) will a Hadoop-based solution really outperform the current ETL performance? If no, what are risks and how can they be tackled? (iii) will on-premise ETL data moved to Hadoop platform be threatened by security attacks? and (iv) what countermeasures can be applied to avoid these obstacles? Based on CM-KMS datasets, for goals *(i) improved availability*, *(ii) improved customizability*, and *(iii) improved interoperability,* a list of obstacles presented to the architect. For example, the potential obstacles *service transient fault* against *improved availability* is retrieved from the dataset. That is, the evidential data in the knowledge base informs that this obstacle is likely to occur due to the performance variability of cloud services or network latency between the Hadoop cloud service and on premise network. CM-KMS datasets, additionally, notify the architect of obstacles against the goal *improved customizability*. These include *un-customizable scalability*, *incomplete APIs*, *inflexible pricing model*, and *proprietary APIs*. To save goal *improved interoperability*, CM-KMS retrieves from the datasets the evidential resolution tactics, namely, *replicate system components*, *adapt data*, and *develop adaptor/wrapper*. For each quality goal, the architect reviews the proposed obstacles by CM-KMS datasets. She shortlists probable ones related to the reengineering scenario. Whether an obstacle is perceived important or not is mainly domain-specific and is based on architect's preference, user experience, statistics about legacy system performance, and available accounts about cloud services.

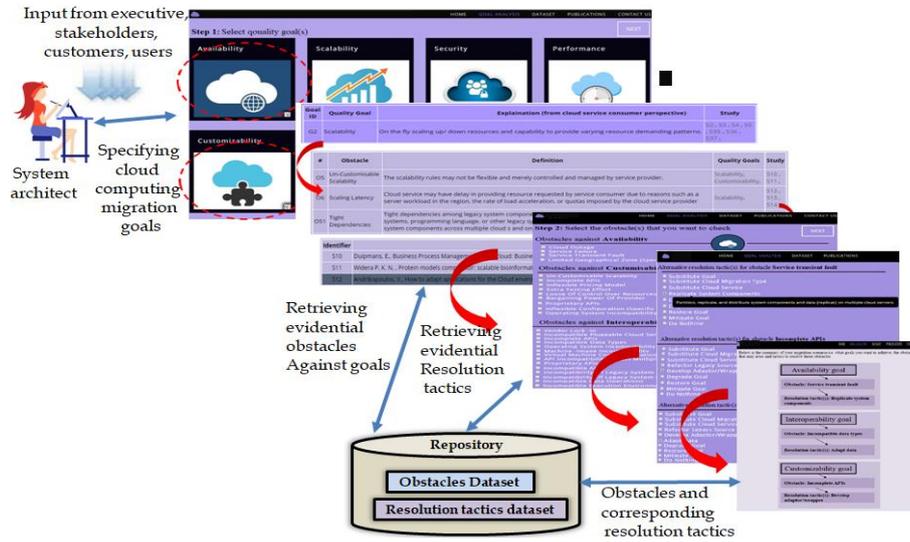

Fig. 3. Analysis steps in CM-KMS - (a) system architect specifies intended cloud migration goals. (b) next, CM-KMS identifies potential obstacles against these goals and alternative resolution tactics based on evidential knowledge the datasets

## 5   Interview evaluation and findings

To analyze the challenges of CM-KMS design and development, we conducted the qualitative method of semi-structured interviews [30]. The interview allowed us to explore the individual challenges stated by the participants in relation to adoption of KMS. Moreover, the interview enabled dynamic interaction based on our experts' background and their responses. Our industry partners helped us recruit a *purposive sample* [29] of interviewees as users with hands-on experience in cloud migration. We asked 11 users (labeled U1-U11 in Table 4) that we were in contact with them during this project to test and perform a peer review of CM-KMS artefact to appraise its usefulness for architects to conduct the early-stage cloud migration requirements analysis in their own project context. The expertise variation of users as shown in Table 4 ranged from software developers to architects, project managers, and academics with years of experience between 2 and 7 in the cloud migration field. The evaluation episode took two months. The interview questions, defined based on DPs, were used to evaluate CM-KMS. The questionnaire included structured interview questions about CM-KMS. The individual interview took between 40-60 minutes (excluding time for the following-up). We used the Qualtrics online survey creator to design and document both interviews and use cases that resulted in 8 pages of comments.

Table 4. Profile of users (PM: Project Manager, SD: Software Developer, SA: System Architect, AR: Academic Researcher

| Expert | Role(s) | Industry sector(s) | Years of exp. | Country |
|---|---|---|---|---|
| U1 | PM | Higher education | 2 | Australia |
| U2 | SA | Utilities | 6 | Saudi Arabia |
| U3 | PM, SA, SD | Retail, Automobile | 7 | Japan |
| U4 | PM | Geo-spatial | 6 | Australia |
| U5 | SA | Healthcare | 3 | Australia |
| U6 | SA | Finance | 5 | Australia |
| U7 | SA | Tourism | 3 | Spain |
| U8 | SA, AR | Utilities | 6 | Switzerland |
| U9 | SA, AR | Higher Education | 4 | Germany |
| U10 | AR | Higher Education | 3 | Germany |
| U11 | AR | Higher Education | 5 | Austria |

All users supported the usefulness and the value of CM-KMS. For example, U1, a project manager with two years of experience in cloud migration stated that "*This repository [CM-KMS] provides a number of considerations that would be useful to developers...*". The adherence of the artefact to DP1, i.e., capturing complete and most relevant evidential knowledge, was acknowledged by words such as "*the breadth and depth of studies that succinctly summarize cloud goals, obstacles and resolution tactics*" (U1), "*comprehensive dataset as a foundation and step by step selection and (static) report*" (U2), "*a comprehensive perspective for the risk assessment and the selection*" (U3), and "*well compiled datasets*", "*sound flow from goal and obstacle to resolution tactics*" (U8). On the other hand, in the view of DPs, we identified a set of nine challenges related to CM-KMS which are described in the following.

**(i) Continuous changes and updates.** Cloud providers constantly improve the performance of their offered services. Thus, some obstacles that are critical now will be ineffective and concerned less in the future: "*cloud providers such as AWS are introducing new services each week and even it's difficult for us to keep [our knowledge base] up with their fast pace*", U6 said. Hence, adherence to DP1 can be a costly investment: "*knowledge base for cloud is useful if populated from whole literature and kept fresh (which usually is a challenge for all knowledge management initiatives that are executed as projects rather than programs*", noted by E8. The participants also were vary in their expected details that needed to be provided by CM-KMS. For instance, an area of concern by U3 was to enhance the datasets inclusivity to cover obstacles against cloud migration goals such as: "*(i) system availability—slow recovery speed from failure (in Aug.2019, by the whole zone failure of AWS Tokyo region, many Japanese services using AWS, lost redundancy and down 12 hours) (ii) system security— lack of* *confidentiality as clients sometimes request the leased line and dedicated hosts where not all public cloud provider can provide these services*, and *(iii) Insufficient cloud service log tracking capability— as clients request the complete log tracking capacity but not all requirements are accepted by cloud providers"*. Likewise, U5 suggested adding obstacles related to "*data protection*" and "*regulations*" in the datasets. A general and long-known problem with KMS, which was confirmed in our project, is the maintenance problem [30] where CM-KMS to be routinely monitored to figure out what enhancements to the datasets should be made and more importantly how to motivate software teams to keep CM-KMS update and contribute their knowledge.

**(ii) Integration of knowledge.** Different CM-KMS, for the most part, are developed and operated by heterogeneous technologies in a software development company working on cloud migration projects. An important component of CM-KMS maintenance is its integrity with other CM-KMS. According to the interviews, CM-KMS should enable them to link and tap into different pre-existing code sharing repositories based on their inquiries. For example, U5 mentioned "*GitHub code sharing platform enables cloud developers to share open-sources source codes and build-in libraries with other developers. This can be used for sharing cloud migration knowledge to connect requirements design and coding*". An implication of this opinion is that the connectivity among CM-KMSs should be ensured and improved on a regular basis as software teams expect to see a holistic rather than a fragmented picture of cloud migration knowledge.

**(iii) Knowledge granularity.** In accordance with determining the level of granularity, i.e. the division of knowledge in terms of size, decomposability, and the extent to which a piece of knowledge is intended to be used as a part of a larger one and stored in CM-KMS [30], the interviews showed that they were not entirely convinced about the level of details provided by our CM-KMS. For example, U2 stated that "*[...] more detailed would be needed… concrete examples of the resolution tactics in action (code snippets, stories, links to presentations)*". We found that the decision on the level of knowledge fragment granularity is a challenging exercise in a CM-KMS for both knowledge consumers as well as for knowledge creators.

**(iv) Preservation of context.** The informants from the interview perceived the importance of the contextual and situational information on actual pitfalls and the logic therein is pivotal in a CM-KMS. An obstacle against a system quality goal might be concerned in one scenario but it has no bearing on another. For example, a goal, e.g. system availability running in the cloud, might be expected to be satisfied 99% of the time and under specific conditions. For such a situation, there should be conditional risks that the architect wants to explore but that are not captured in CM-KMS. We annotated each obstacle regarding its associated cloud service delivery models such as IaaS (infrastructure as a service), SaaS (software as a service), PaaS (Platform as a service) proposed by [1] indicating the relationship between the obstacle occurrence with different service delivery models. This illustrates the following statement: "*there are varying perspectives in a cloud migration: network, applications, operations, databases, project manager, architect and executive. Knowledge management repository should be tailored to these perspectives, making it more relatable to each party*", U1 mentioned. The participants had division opinions about the enhancements of current CM-KMS such as "*having some case studies or including practical examples*" U3 and "*concrete examples of the resolution tactics in action, e.g., code snippets, stories, links to presentations*", U8. It was turned out that CM-KMS lacks a classification of "*obstacles and tactics based on whether they are intrinsically related to public, private cloud*", mentioned by U4.

**(v) Automation.** If a CM-KMS can automatically discern obstacles based on cloud migration project information, e.g. legacy system architecture, and implement recommendations or workflow to negate them at the requirements level, then it will be helpful tool for architects and software teams. The automatic uncovering of such complex logics in migration processes to cloud can give the theoretical importance that can be accentuated and delivered through a typical CM-KMS design. Nevertheless, if the system merely informs potential obstacles and resolution tactics for its users, the aspect of intriguing and theoretically important design of CM-KMS is lost or low.

**(vi) Deconstruction of traditional knowledge.** The cloud computing technology leverages concepts and principles of distributed architecture design and been evolved out of former technologies like grid computing with characteristics such as computing power on demand, reducing the cost of computing, and flexible transforming computers. For instance, the obstacle *service transient fault* proposed in our knowledge base occurs similarly in the grid computing architecture, nevertheless, it reappears taking to an extreme level in cloud computing as systems in cloud platforms are operating at a massive scale. One would expect that CM-KMS would have drawn on the existing bodies of knowledge regarding the conventional legacy system reengineering to a new platform, where evidential knowledge has been attained and tested thoroughly over time. An implication of this findings has been the inclusion of 36 obstacles associated with the common system quality goals in our CM-KMS datasets.

**(vii) Dependency on experts.** The actual root causes for certain cloud migration risks and countermeasures might be highly personal, subjective, and communication-related and cannot be explicitly tackled with simple guidelines and decision support. Like many software engineering projects, cloud migration is complex beasts and deal not only with the commonly known software engineering practices and capabilities of the customer organizations but also with somewhat tricky co-creative practices of the stakeholders and software teams involved that might be impractical to document. Moreover, confirmatory to [14], our findings showed that architects may also rely on other techniques such as *learning by doing* or *individual mental models* to unveil new obstacles and resolutions tactics that are not explicitly reported in the literature and hence reflected in CM-KMS datasets.

**(viii) Hybrid knowledge of both vendor-specific and vendor-neutral cloud platforms.** To increase applicability, the variety of public or private cloud services, each with inherent features, entails exploiting both general and special cases of knowledge fragments for inclusion in a CM-KMS. U2 stated that "*the repository would be more interesting if it could incorporate cloud platform-specific knowledge of popular cloud platforms, like AWS, Azure. Some of the obstacles might only apply to specific platforms or the extent of the obstacle is different on different cloud platforms*". This led us to iterate the design phase of DSR where we extended CM-KMS datasets by adding 15 identified evidential data items from published reports by cloud vendors Amazon,

Google, and Rackspace. U4 suggested the classification of the datasets based on service providers, i.e. "*I believe, specifically classified knowledge (e.g., public and private cloud) could further enhance the usefulness*". This challenge again impose the cost of KMS development and maintenance.

**(ix) Parsimony.** Empirical reports in the cloud migration literature may not always elaborate on their findings, project context, and lessons learned. Many cloud architects and organizations do not share all experienced issues of transition to cloud with outsiders because of privacy, regulatory issues, and intellectual property matters. This may affect the reliability of a proposed CM-KMS for cloud migration requirements analysis and decision making. Like many software projects, legacy-to-cloud may produce tacit knowledge, much of which are not explicitly documented, publically available, and even difficult to extract from the minds of architects.

# 6 Conclusion

With one sample scenario of CM-KMS application and 11 interview participants, our research has a relatively small sample size. To lessen this effect, we strived to identify and interview participants with hands-on real-world experience in migrating legacy software applications to cloud platforms in large companies or well-recognized research in this field. Nevertheless, we do not claim and guarantee that our findings, i.e. HMCS project case and interview results, are generally applicable to other projects and other CM-KMS design endeavours. Future studies are required to further confirm our results.

We are fully aware that multiple KMS projects that resulted in knowledge sharing repositories have not been eventually adopted. This is partially due to similar challenges that we reported in this paper. Additionally, actual processes involve so many personal, interpersonal, organizational, inter-organizational, environmental, and situational factors while knowledge repositories may provide only superficial help for software teams. There are fruitful directions that we can take in further. We have populated a larger number of evidential data in our datasets. To improve the adherence to DP5, we plan to embark these datasets to public GitHub platform to enable practitioners and researchers in cloud computing community to add their own empirical findings to enrich the datasets. Currently, CM-KMS design has tended to be as lightweight as possible to minimize the process workload of requirements and decision analysis. Nevertheless, the representation of the datasets can be further enhanced regarding DP7. For example, the notion of resolution tactic/countermeasure is informally explained in the datasets, which makes it difficult to query them to find how a particular risk can be handled. Augmenting the specification of knowledge fragments in the datasets by a particular ontology will assist the knowledge base searches. Finally, cloud computing technology provides a backbone to support next generation of computing based systems such as Internet of Things, Edge computing, and Blockchain, which analogously raise new requirements. Future research can extend the presented KMS in this research to gather and package new evidential data specific to these modern Internet-based computing technologies. One can enhance this track research with a focus on KMS for emerging trends like migration to micro-services and Internet of Things.

In terms of the practical applicability, the identified challenges in this study helps cloud architects to understand the potential failure challenges in development and adoption of KMS for cloud migration projects and avoid them in line with the proposed DPs. In large IT-based organizations with data-intensive and critical legacy systems and a large number of cloud architects and developers, the main concern should be creating such motivations that justify capturing, sharing, and updating KMS to utilize others' cloud migration knowledge and experiences. On the other hand, for cloud architects in small IT-based organizations, because of resource constraints, they should make an informed decision in assessing if ever-evolving and costly KMS is needed to support cloud migration process and establish appropriate fit via using their available resources efficiently. For researchers, the eight design principles and nine challenges related to designing cloud specific KMS can be a reliable reference for future research. For example, researchers can focus on each challenge and propose practical countermeasure to mitigate the challenge in line design principles.